\documentclass[9pt,twocolumn,twoside]{osajnl}

\usepackage{csquotes}
\usepackage{amsmath}
\usepackage{amssymb}
\usepackage{graphics}
\usepackage{graphicx}
\usepackage{epstopdf}
\usepackage{dcolumn}
\usepackage{bm}
\usepackage{color}

\newcommand{\beq}{\begin{equation}}
\newcommand{\eeq}{\end{equation}}
\newcommand{\bdm}{\begin{displaymath}}
\newcommand{\edm}{\end{displaymath}}
\newcommand{\bary}{\begin{array}}
\newcommand{\eary}{\end{array}}
\newcommand{\beqary}{\begin{eqnarray}}
\newcommand{\eeqary}{\end{eqnarray}}
\newcommand{\bize}{\begin{itemize}}
\newcommand{\eize}{\end{itemize}}
\newcommand{\bc}{\begin{center}}
\newcommand{\ec}{\end{center}}
\newcommand{\bfig}{\begin{figure}}
\newcommand{\efig}{\end{figure}}
\newcommand{\bd}{\begin{document}}
\newcommand{\ed}{\end{document}}

\newcommand{\mtA}{\mathcal{A}}
\newcommand{\mtE}{\mathcal{E}}
\newcommand{\mtI}{\mathcal{I}}
\newcommand{\mtP}{\mathcal{P}}
\newcommand{\mtN}{\mathcal{N}}
\newcommand{\mtR}{\mathcal{R}}
\newcommand{\Ba}{{\bf a}}
\newcommand{\Be}{{\bf e}}
\newcommand{\Bf}{{\bf f}}
\newcommand{\Bg}{{\bf g}}
\newcommand{\Bk}{{\bf k}}
\newcommand{\Bl}{{\bf l}}
\newcommand{\Bn}{{\bf n}}
\newcommand{\Bp}{{\bf p}}
\newcommand{\Bq}{{\bf q}}
\newcommand{\Br}{{\bf r}}
\newcommand{\Bs}{{\bf s}}
\newcommand{\Bx}{{\bf x}}

\newcommand{\BE}{{\bf E}}
\newcommand{\BD}{{\bf D}}
\newcommand{\BG}{{\bf G}}
\newcommand{\BK}{{\bf K}}
\newcommand{\BQ}{{\bf Q}}
\newcommand{\BR}{{\bf R}}
\newcommand{\BV}{{\bf V}}
\newcommand{\BU}{{\bf U}}
\newcommand{\BW}{{\bf W}}

\newcommand{\Bone}{{\bf 1}}

\newcommand{\lgl}{{\langle}}
\newcommand{\rgl}{{\rangle}}
\newcommand{\hx}{{\hat{x}}}
\newcommand{\hz}{{\hat{z}}}
\newcommand{\hy}{{\hat{y}}}
\newcommand{\hp}{{\hat{p}}}
\newcommand{\ha}{{\hat{a}}}
\newcommand{\hA}{{\hat{A}}}
\newcommand{\hB}{{\hat{B}}}
\newcommand{\hF}{{\hat{F}}}
\newcommand{\hL}{{\hat{L}}}
\newcommand{\hS}{{\hat{S}}}
\newcommand{\hD}{{\hat{D}}}
\newcommand{\hs}{{\bf \hat{s}}}
\newcommand{\Bkk}{{\bf\hat{k}}}
\newcommand{\bspp}{{\bf s_{\perp}}}
\newcommand{\hro}{{\hat{\rho}}}

\newcommand{\specbf}[1]{\mbox{\boldmath $\bf#1$}}
\newcommand{\BmtE}{\specbf{\mathcal{E}}}
\newcommand{\BmtP}{\specbf{\mathcal{P}}}
\newcommand{\bepsilon}{\specbf{\epsilon}}
\newcommand{\Brho}{\specbf{\rho}}
\newcommand{\Bkappa}{\specbf{\kappa}}
\newcommand{\Btheta}{\specbf{\theta}}
\newcommand{\Bxi}{\specbf{\xi}}
\newcommand{\Bkap}{\specbf{\kappa}}
\newcommand{\BGam}{\specbf{\Gamma}}
\newcommand{\Bsig}{\specbf{\sigma}}
\newcommand{\BLam}{\specbf{\Lambda}}
\newcommand{\specme}[1]{\mbox{$#1$}}
\newcommand{\epsone}{\specme{\epsilon_{\parallel}}}
\newcommand{\epstwo}{\specme{\epsilon_{\perp}}}

\newcommand{\pd}[2]{\frac{\partial #1}{\partial #2}}%
\newcommand{\ave}[1]{\langle #1 \rangle}%
\newcommand{\half}{\textstyle \frac{1}{2}}%
\newcommand{\ihalf}{\textstyle \frac{i}{2}}%
\newcommand{\third}{\textstyle \frac{1}{3}}%
\newcommand{\six}{\textstyle \frac{1}{6}}%
\journal{ol}

\setboolean{shortarticle}{true}

\title{Airy beams on incoherent background}

\author[1]{Morteza Hajati}
\author[1]{Vincent Sieben}
\author[1,2,3]{Sergey A. Ponomarenko}

\affil[1]{Department of Electrical and Computer Engineering, Dalhousie University, Halifax, Nova Scotia, B3J 2X4, Canada}
\affil[2]{Department of Physics and Atmospheric Science, Dalhousie University, Halifax, Nova Scotia, B3H 4R2, Canada}
\affil[3]{serpo@dal.ca}

\dates{Compiled \today}

\doi{\url{http://dx.doi.org/10.1364/OL.XX.XXXXXX}}

\begin{abstract}
We present a class of diffraction-free partially coherent beams each member of which is comprised of a finite-power, non-accelerating Airy bump residing on a statistically homogeneous, Gaussian-correlated background. We examine free-space propagation of soft apertured realizations of the proposed beams and show that their evolution is governed by two spatial scales: the coherence width of the background and aperture size. A relative magnitude of these factors determines the practical range of propagation distances over which the novel beams can withstand diffraction. The proposed beams can find applications to imaging and optical communications through random media.
\end{abstract}

\setboolean{displaycopyright}{true}

\begin{document}
\maketitle
Partially coherent light sources have attracted a great deal of interest over the years because they generate a plethora of structured statistical light fields with intriguing properties~\cite{Cai17,Chen20a} that carry promise for multiple applications~\cite{Korotkova20a}. The majority of partially coherent optical fields studied to date, though, have been statistically homogeneous~\cite{MW}. Yet, statistically nonuniform fields have also been explored, both theoretically~\cite{PSA01,PSA07,Gori09,Laj11,Chen19} and experimentally~\cite{Bog03,Ostr17,OK18,Zhu19}. In particular, propagation invariant random fields have been shown theoretically~\cite{PSA07,Gori09} and verified experimentally~\cite{Zhu19} to necessarily possess nonuniform correlations as they manifest themselves as bumps or dips residing atop of a statistically homogeneous background. Such statistical diffraction free beams have been shown to possess universal self-similar asymptotic propagation properties in random media~\cite{PSA20}. Several generalizations of random diffraction free beams were proposed~\cite{Rady,Hyde19,Hyde20}, including, notably, white-light (polychromatic) dark/antidark fields on incoherent background~\cite{Rady}.

At the same time, coherent diffraction free Airy waves in general~\cite{Berry} and optical Airy beams in particular~\cite{Chris1} have attracted much attention since their experimental realization~\cite{Chris2}, not the least bit because of their intriguing accelerating nature and a wealth of potential applications~\cite{Airy-rev}. Partially coherent Airy beams have also been realized~\cite{Seg}. However, the latter, albeit accelerating, are not genuine diffraction free beams, even in the ideal, infinite-power limit since their cross-spectral density is not of the form of a bump (or dip) on a statistically uniform background. This observation begs a fundamental question: Can one generate bona fide diffraction free random Airy beams, and if so, will their finite-power realizations accelerate in free space?

In this Letter, we demonstrate how a family of genuine diffraction free, partially coherent Airy sources can be constructed. We show that any member of the family is represented by an Airy bump situated on a Gaussian correlated uniform background. The coherence properties of any beam generated by such a source are controlled by a single dimensionless parameter, a correlation parameter given by the ratio of a coherence width of the background to a characteristic width of the bump. We study free-space propagation of Gaussian apertured realizations of the proposed beams and show that while their acceleration is suppressed, the distance over which the new beams can withstand diffraction is determined by the relative magnitude of two spatial scales: the background coherence width and aperture size. 

We start with a general representation for the cross-spectral density of any nonnegative definite 1D partially coherent source in the form~\cite{Gori07}
	\beq\label{Gori}
		W(x_1,x_2)=\int_{-\infty}^{\infty} dk\,p(k) \mtA^{\ast}(k,x_1)\mtA(k,x_2),
			\eeq
where we dub the nonnegative real function $p(k)\geq 0$, $p^{\ast}(k)=p(k)$, a spectral distribution and $\{\mtA(k,x)\}$ generalized modes of the source corresponding to a continuous spectrum labelled by the variable $k$.

Let us consider a Gaussian spectral distribution of the modes,
	\beq\label{p}
		p(k)=\frac{\sigma_c}{\sqrt{2\pi}}e^{-k^2\sigma_c^2/2},
			\eeq
where $\sigma_c$ is the source coherence length, and the plane-wave modes with cubic phase chirp, so that
	\beq\label{A}
		\mtA(k,x)=\sqrt{2I_0}\cos\left(\six \sigma_I^3 k^3 +\half kx\right).
			\eeq
$I_0$ and $\sigma_I$, entering Eq.~(\ref{A}), set the scales of the source intensity and spatial width of the bump, respectively. Numerical coefficients are chosen for convenience. In physics terms, Eqs.~(\ref{Gori}) through~(\ref{A}) imply that the source is composed of a continuum of phase chirped plane waves, with each plane wave contribution to the overall source intensity being weighed with the Gaussian in Eq.~(\ref{p}). We can also surmise that $\sigma_c$ controls the source state of coherence. In particular, in the limit $\sigma_c \rightarrow \infty$, we can infer at once that $p(k)\rightarrow \delta(k)$, implying the source engenders a chirped plane wave in the fully coherent limit. As we will see in more precise terms shortly, $\sigma_c$ plays the role of the background coherence width. 

It follows from the trigonometric identity,
	\beq
		\cos\alpha\cos\beta=\half[\cos(\alpha-\beta)+\cos(\alpha+\beta)],
			\eeq
that
	\beq\label{aux1}
		\mtA^{\ast}(k,x_1)\mtA(k,x_2)=I_0 [\cos (kx_{-}/2) +\cos(\sigma_I^3 k^3/3+kx_{+})],
			\eeq
where we introduced the center-of-mass and difference coordinates as
	\beq\label{xX}
		x_{+}=(x_1+x_2)/2, \hspace{0.5cm} x_{-}=x_1-x_2.
			\eeq
On substituting from Eq.~(\ref{aux1}) into Eq.~(\ref{Gori}), performing Gaussian integrations and making use of the integral representation of the Airy function,
	\beq
		\mathrm{Ai}(x/\sigma_I)=\frac{\sigma_I}{2\pi}\int_{-\infty}^{\infty}ds\,\exp\left(\frac{i\sigma_I^3 s^3}{3}+isx\right),
			\eeq
we arrive at the cross-spectral density of an Airy beam on incoherent background (ABIB) in the form
\beq\label{Wdfd}
		\overline{W}_{\mathrm{A}} (X_{-},X_{+})=\overline{W}_{\mathrm{bg}}( X_{-})+\overline{W}_{\mathrm{bp}}(X_{+}),
			\eeq
where
	\beq\label{bg}
		\overline{W}_{\mathrm{bg}}( X_{-})= \exp\left(-\frac{X_{-}^2}{8\xi_c}\right),
			\eeq
is the cross-spectral density of a statistically uniform, Gaussian-correlated background and
	\beq\label{bp}
		\overline{W}_{\mathrm{bp}}(X_{+})=P_{\mathrm{1D}} e^{\xi_c^3/12} e^{\xi_c X_{+}/2}\mathrm{Ai}(X_{+}-X_0)
			\eeq
is that of an Airy bump atop of the background. Here we introduced dimensionless variables, $\overline{W}=W/I_0$ and $X_{\pm}=x_{\pm}/\sigma_I$. Further, we introduced a correlation parameter $\xi_c=\sigma_c^2/\sigma_I^2$ as well as a coordinate $X_0=-\xi_c^2/4$ of the Airy function maximum and a total power $P_{\mathrm{1D}}=\sqrt{2\pi\xi_c}$ of the Airy bump. The source intensity $\overline{I}_{\mathrm{A}}(X)=\overline{W}_{\mathrm{A}}(X,X)$, can be determined at once from Eq.~(\ref{Wdfd}) by setting $X_{-}=0$ and $X_{+}=X$ which yields the expression
\beq\label{I1d}
		\overline{I}_{\mathrm{A}}(X)=1+P_{\mathrm{1D}} e^{\xi_c^3/12} e^{\xi_c X/2}\mathrm{Ai}(X-X_0).	
			\eeq

At this point, let us make some instructive observations. First, Eqs.~(\ref{Wdfd}) through~(\ref{I1d})  imply that the newly constructed sources generate bona fide diffraction-free beams as the source cross-spectral density has the propagation invariant form that was first explicitly derived  in~\cite{PSA07}. Second, the properties of a family of ABIBs are governed---in dimensionless units---by the single correlation parameter $\xi_c$. Third, ideal Airy beams on incoherent background do not accelerate: the peak intensity of the bump is located at a fixed position dependent on the background coherence. Fourth, each Airy bump carries a finite power $P_{\mathrm{1D}}$ which increases with the coherence level of the background. The last two properties of the discovered beams bring into focus sharp contrast between ABIBs and fully coherent Airy beams. Interestingly, the exponential cut-off of the source intensity at negative coordinates arises naturally in the partially coherent case. 

\begin{figure}[h!]
\centering
   \includegraphics[width=3.5in]{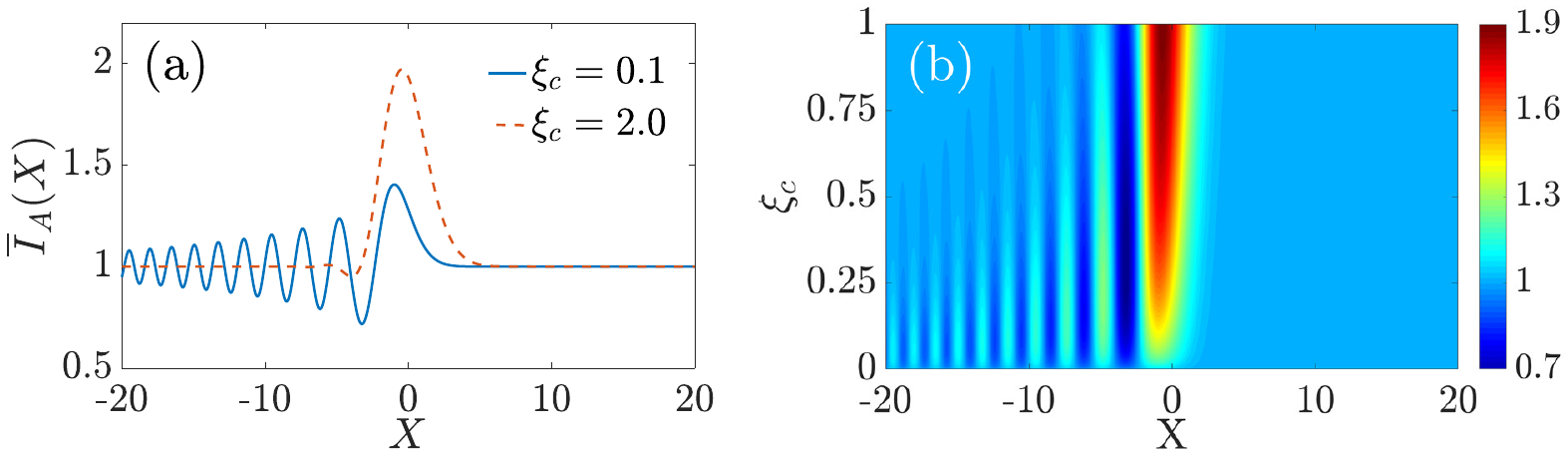}
  \caption{(a) Intensity profile of an ideal ABIB as a function of $X$ for two values of the correlation parameter $\xi_c$: $\xi_c=0.1$ (solid curve) and $\xi_c=2$ (dashed curve). (b) Contour plot of the ABIB intensity. }
\label{fig1}
\end{figure}
To visualize the spatial structure of a typical ABIB, we display the intensity distribution of the beam for two values of the correlation parameter in Fig.~\ref{fig1}. We can infer from the figure that while a rather correlated ABIB, such that $\sigma_I\leq \sigma_c$ corresponding to $\xi_c\geq 1$, has a relatively high-intensity bump, a nearly uncorrelated ABIB with $\sigma_c \ll \sigma_I$ features a short bump and a long oscillatory left tail. This behavior can be explained by noticing that a weakly correlated background breaks up the beam into a number of essentially uncorrelated beamlets, each having nearly the same peak intensity as its neighbor. While the oscillations (beamlets) in the left tail of the ABIB are gently cut-off at long $X$ by the exponential factor $e^{\xi_c X/2}$, the ones in the right tail are sharply suppressed by the super-exponential asymptotics of the Airy function, $\mathrm{Ai}(X)\sim e^{-(2/3)X^{3/2}}$ as $X\rightarrow +\infty$. 

\begin{figure}[h!]
\centering
   \includegraphics[width=3.5in]{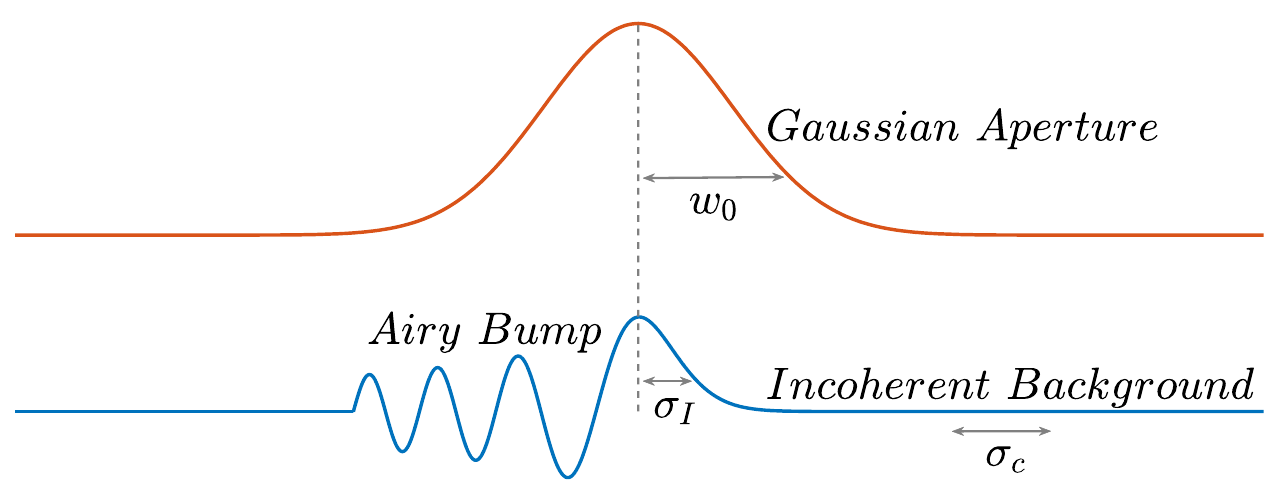}
  \caption{Illustrating the hierarchy of transverse scales associated with a Gaussian apertured Airy beam on incoherent background.}
\label{fig2}
\end{figure}

An ideal ABIB carries infinite power because it dwells on a statistically uniform background, and hence is not realizable in the laboratory. We are then bound to consider a ``softly" apertured ABIB, which can be experimentally realized with the aid of a Gaussian amplitude mask, for example, and examine free-space paraxial propagation of the Gaussian apertured ABIB. We assume that the aperture width $w_0$ dwarfs the transverse width of the bump, $\sigma_I\ll w_0$. We schematically illustrate the hierarchy of transverse scales governing free-space propagation of finite-power ABIBs in Fig.~\ref{fig2}. The cross-spectral density of the apertured ABIB source has the form
	\beq\label{Ws}
		W^{(0)}(x_1,x_2)=\exp\left(-\frac{x_1^2+x_2^2}{2w_0^2}\right)\,W_{\mathrm{A}}(x_1,x_2),
			\eeq
where $W_{\mathrm{A}}$ is the cross-spectral density of the ideal ABIB given by Eqs.~(\ref{Wdfd}) through~(\ref{bp}). The intensity of the apertured ABIB with the carrier wavenumber $k_0$ in any cross-section $z\geq 0$ is given by a Fresnel transform~\cite{MW}
	\begin{align}\label{Iz}
		I(x,z)&=\left(\frac{k_0}{2\pi z}\right)\int dx_1 \int dx_2\,W^{(0)}(x_1,x_2) e^{ik_0 (x-x_2)^2/2z} \nonumber \\
			&\times e^{-ik_0 (x-x_1)^2/2z}.
			\end{align}
\begin{figure}[t]
\centering
   \includegraphics[width=3.5in]{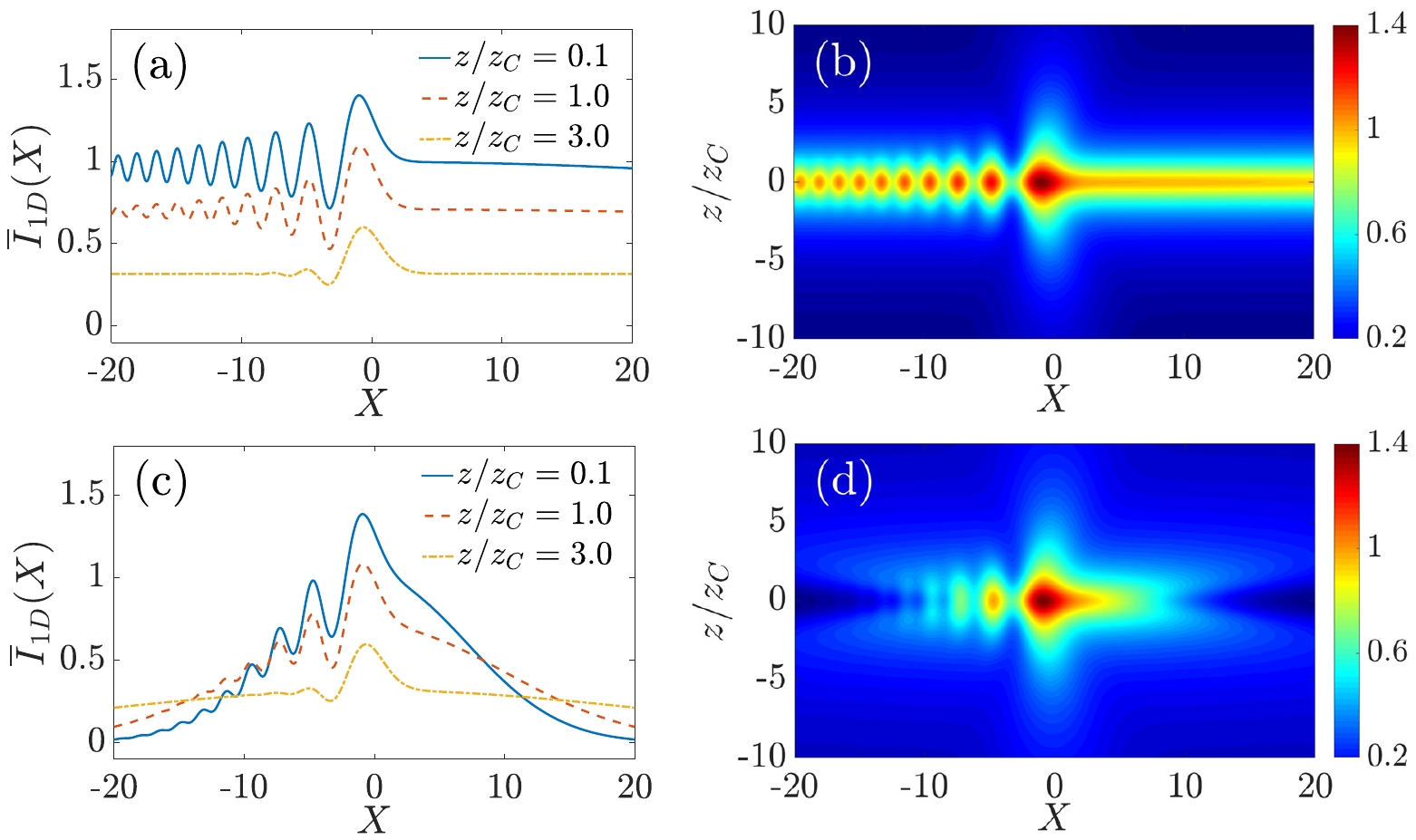}
  \caption{(a) and (c) Intensity profiles of a nearly uncorrelated Gaussian apertured ABIB as a function of $X$ for three propagation distances: $z=0.1 z_c$ (solid curve), $z=z_c$ (dashed curve), and $z=3z_c$ (dash-dotted curve) and the aperture size (a) $\epsilon=0.01$ and (c) $\epsilon=0.1$. (b) and (d) Contour plots of the Gaussian apertured ABIB with (b) $\epsilon=0.01$ and (d) $\epsilon=0.1$. The correlation parameter is $\xi_c=0.1$. }
\label{fig3}
\end{figure}			
On substituting from  Eqs.~(\ref{Wdfd}) through~(\ref{bp}) as well as from Eq.~(\ref{Ws}) into~(\ref{Iz}) and converting to dimensionless center-of-mass and difference coordinates, we obtain, after quite tedious but straightforward integrations involving Gaussians and the Airy function representation, the following expression
	\beq\label{Iaz}
		\overline{I}(X,z)=\overline{I}_{\mathrm{bg}}( X,z)+\overline{I}_{\mathrm{bp}}(X,z),
			\eeq
with
	\beq\label{bgz}
		\overline{I}_{\mathrm{bg}}( X,z)=\frac{1}{\sqrt{1+z^2/z_c^2}}\,\exp\left(-\frac{\epsilon^2 X^2}{1+z^2/z_c^2}\right),
			\eeq
and
	\begin{align}\label{bpz}
		\overline{I}_{\mathrm{bp}}(X,z)&=\overline{I}_0(z)\exp\left[\frac{\xi_c X}{2}\frac{(1+z^2/z_c^2)}{ (1+z^2/z_R^2)^2}\right] \nonumber \\
			&\times  \exp\left(-\frac{\epsilon^2 X^2}{1+z^2/z_R^2}\right) \mathrm{Ai}\left[\frac{X-X_0(z)}{1+z^2/z_R^2}\right].
			\end{align}
In Eqs.~(\ref{Iaz}), ~(\ref{bgz}) and~(\ref{bpz}) we introduced a small dimensionless parameter $\epsilon=\sigma_I/w_0\ll 1$ and 	
\beq\label{I0}
				\overline{I}_0(z)=\frac{P_{\mathrm{1D}}}{\sqrt{1+z^2/z_R^2}}\exp\left[\frac{\xi_c^3}{12}\left(\frac{1+z^2/z_c^2}{1+z^2/z_R^2}\right)^3\right],
					\eeq
as well as
		\beq\label{X0z}
			X_0 (z)=-\frac{\xi_c^2}{4}\frac{(1+z^2/z_c^2)^2}{1+z^2/z_R^2}.
				\eeq
				
The physics of apertured ABIB evolution in free space is incapsulated in two longitudinal scales: a Rayleigh range $z_R=k_0 w_0^2$ of the embedding Gaussian beam as well as a characteristic diffraction length $z_c=\sqrt{2}k_0 \sigma_c w_0$ associated with finite background coherence. We can infer from Eq.~(\ref{bgz}) that the Gaussian background spreads, so that its coherence state changes appreciably at the distances of the order of $z_c$. By the same token, it follows from Eqs.~(\ref{bpz}), ~(\ref{I0}) and~(\ref{X0z}) that while the bump maintains its Airy shape, it spreads slowly over a characteristic distance of the order of $z_R$. It can then be viewed as practically diffraction-free over distances $z\ll z_R$. A specific evolution scenario of the ABIB depends on whether the background is nearly incoherent, $\sigma_c\ll\sigma_I$, or highly coherent, $\sigma_I\ll\sigma_c$. In the former case, $z_c$ sets the scale for diffraction, while in the latter, a characteristic distance over which the ABIB succumbs to diffraction is determined by the Rayleigh range of the Gaussian aperture. 

\begin{figure}[t]
\centering
   \includegraphics[width=3.5in]{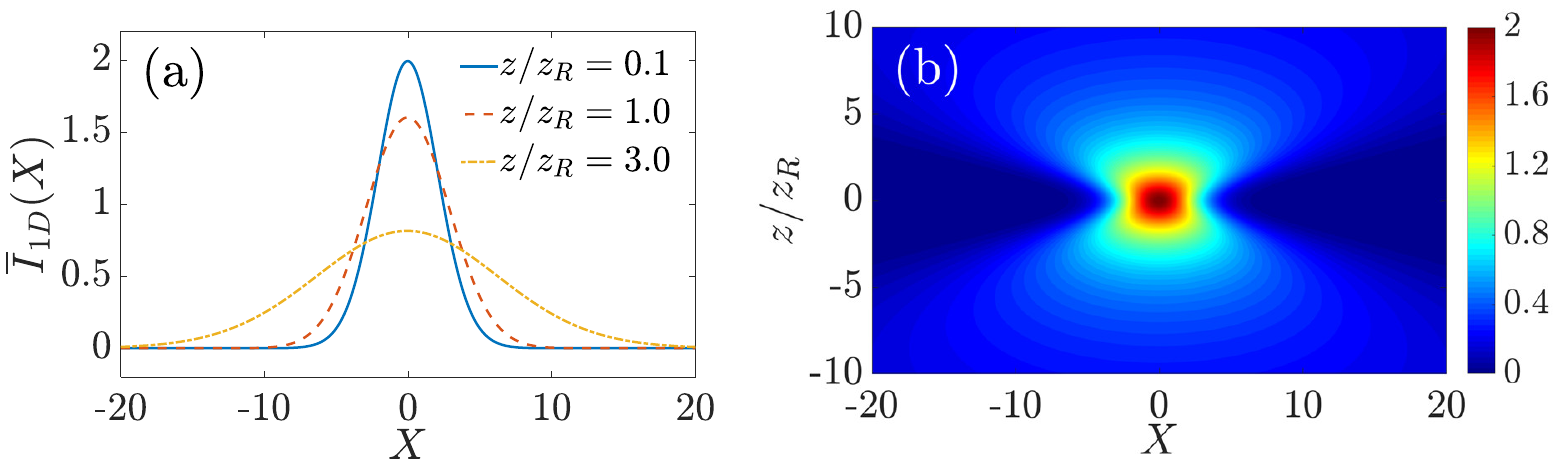}
  \caption{(a) Intensity profiles of a strongly correlated Gaussian apertured ABIB as a function of $X$ for three propagation distances: $z=0.1 z_R$ (solid curve),  $z=z_R$ (dashed curve), and $z=3z_R$ (dash-dotted curve) and the aperture size $\epsilon=0.3$. (b) Contour plot of the Gaussian apertured ABIB. The correlation parameter is $\xi_c=10$. }
\label{fig4}
\end{figure}

To illustrate our qualitative conclusions, we exhibit the evolution of an apertured Airy bump on a nearly incoherent background (weakly correlated ABIB) in Fig.~\ref{fig3} for three propagation distances measured in the units of $z_c$ which sets the scale for diffraction there. We consider two cases: $\epsilon=0.01$ and $\epsilon=0.1$ corresponding to a very large and moderately sized apertures, respectively. In the wide-aperture limit, see Fig~\ref{fig3}(a), the ABIB remains nearly diffraction-free. The beam maintains its ideal structure with a long oscillatory tail, although the peak intensities of the main and secondary lobes diminish with the propagation distance, as does the magnitude of the background intensity. As the aperture size is reduced, though, the ABIB structure undergoes interesting evolution. The beam tails, which are partially cut off by the aperture at short propagation distances from the source plane, reappear on propagation farther away, as the ABIB gradually transfers the power from its main lobe into the tails. This scenario is evident in Fig.~\ref{fig3} (c). Eventually, over long enough propagation distances, the side lobes virtually disappear and the tightly apertured ABIB profile becomes nearly identical to that generated by diffraction of a loosely apertured ABIB. This conclusion can be confirmed by comparing the dash-dotted curves corresponding to the propagation distance $z=3z_c$ in panels (a) and (c) of Fig.~\ref{fig3}. Thus, a nearly universal shape of the weakly correlated ABIB emerges over sufficiently long propagation distances, regardless of the size of the Gaussian aperture. The magnitude of the propagation distance over which the reshaping toward the asymptotic profile occurs does depend on the aperture size through the dependence of $z_c$ on $w_0$, of course.  At the same time, embedding an Airy bump on a fairly coherent background (strongly correlated ABIB) into a Gaussian envelope results in the ABIB spreading over a few characteristic Rayleigh distances as can be seen in Fig.~\ref{fig4}. A  strongly correlated ABIB diffracts much like an ordinary Gaussian beam which makes it less attractive for potential applications than its weakly correlated counterpart. We also notice by inspecting the contour plots in Figs.~\ref{fig4}(b) and~\ref{fig3}(b,d) that the ABIB acceleration is strongly suppressed in the partially coherent case, irrespective of the aperture size and coherence level of the background. 

Finally, a generalization to a 2D ABIB is straightforward. To this end, we can represent the cross-spectral density of the 2D ABIB as
	\beq
		W(\Br_1,\Br_2)=\int d\Bk\,p(\Bk) \mtA^{\ast} (\Bk,\Br_1)\mtA(\Bk,\Br_2),
			\eeq
where 
	\beq\label{p2d}
		p(\Bk)=\prod_{j=x,y}\frac{\sigma_{cj}}{\sqrt{2\pi}}e^{-k_j^2\sigma_{cj}^2/2},
			\eeq
and
	\beq\label{A2d}
		\mtA(\Bk,\Br)=\sqrt{2I_0}\cos\left(\six \sum_{j=x,y}\sigma_{Ij}^3 k_j^3 +\half \Bk\cdot\Br\right).
			\eeq
\begin{figure}[h!]
\centering
   \includegraphics[width=3.5in]{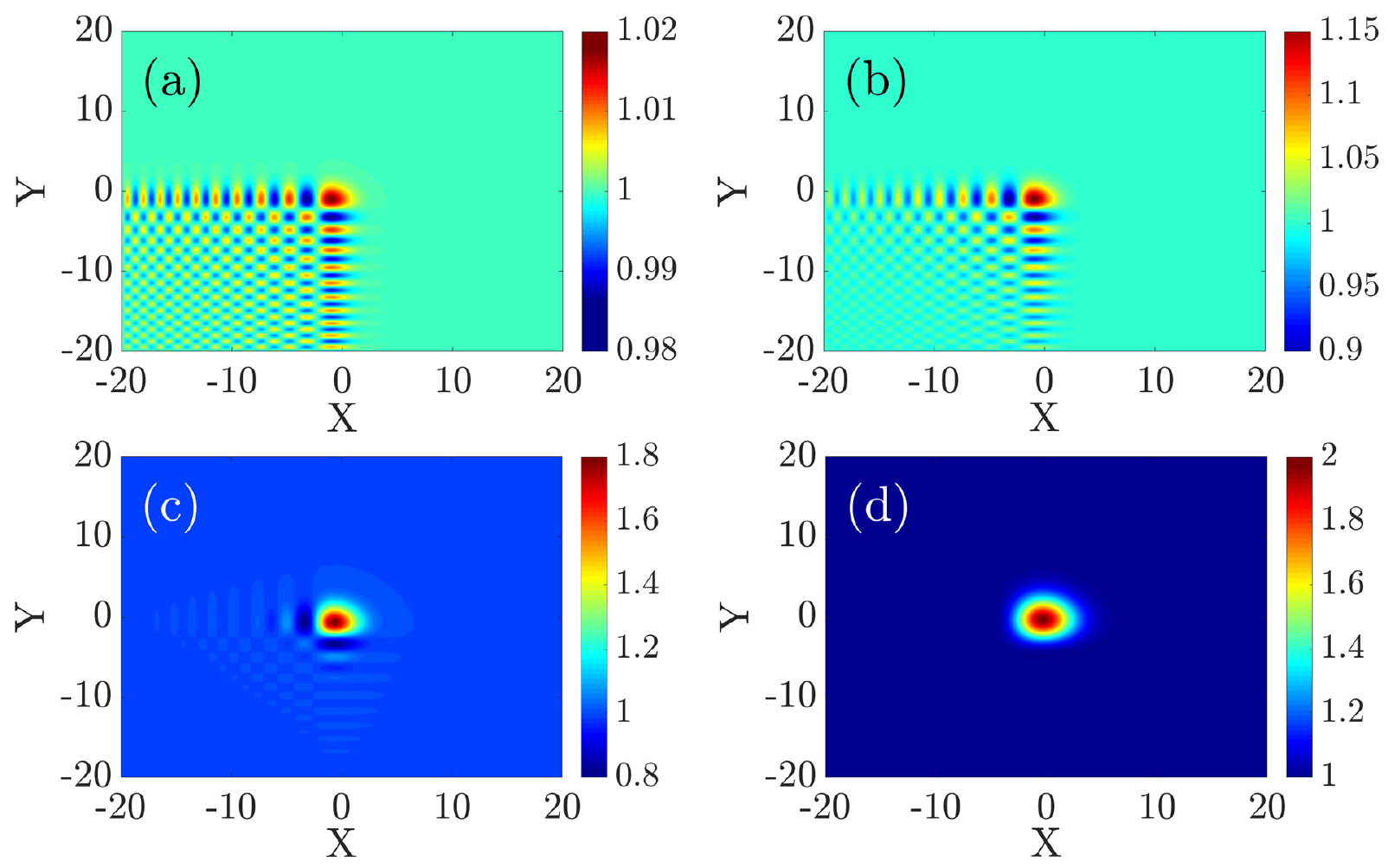}
  \caption{Contour plots of ideal 2D ABIB corresponding to (a) $\xi_c=0.01$, (b) $\xi_c=0.1$, (c) $\xi_c=1$ and (d) $\xi_c=3$.  }
\label{fig5}
\end{figure}
Here we introduced the background coherence  $\sigma_{cj}$ and bump intensity $\sigma_{Ij}$ widths along the corresponding axes $j=x,y$ of the Cartesian coordinate system; the source is, in general, anisotropic.  Repeating exactly the same steps as we took to derive the cross-spectral density of the 1D ABIB, we arrive at the expression
	\beq
		\overline{W}_{\mathrm{A}} (\BR_{-},\BR_{+})=\overline{W}_{\mathrm{bg}}( \BR_{-})+\overline{W}_{\mathrm{bp}}(\BR_{+}),
			\eeq		
with
	\beq
		\overline{W}_{\mathrm{bg}}(\BR_{-})= \exp\left(-\frac{X_{-}^2}{8\xi_{cx}}-\frac{Y_{-}^2}{8\xi_{cy}}\right),
			\eeq
and
	\begin{align}
		\overline{W}_{\mathrm{bp}}(\BR_{+})&=P_{\mathrm{2D}} \exp\left[(\xi_{cx}^3+\xi_{cy}^3)/12\right] e^{\xi_{cx}X_{+}/2}e^{\xi_{cy}Y_{+}/2}
			\nonumber \\
		&\mbox{}\times \mathrm{Ai}(X_{+}-X_0)\mathrm{Ai}(Y_{+}-Y_0),
			\end{align}
where we introduced correlation parameters $\xi_{cj}=\sigma_{cj}^2/\sigma_{Ij}^2$ and a total power of a 2D Airy bump as $P_{\mathrm{2D}}=2\pi\sqrt{\xi_{cx}\xi_{cy}}$. Notice that although the cross-spectral densities of the bump and background individually factorize in Cartesian coordinates, the cross-spectral density of a 2D Airy beam on incoherent background does not.  We exhibit the intensity profiles of ideal 2D ABIBs in Fig.~\ref{fig5}. We can infer from the figure by comparing panels (a) through (d) that as the source becomes progressively more correlated---or the background coherence level increases---the ABIB power localizes more within the main lobe. This picture is consistent with our findings for 1D ABIBs. 

 In summary, we have introduced a class of bona fide diffraction free statistical beams, Airy beams on incoherent background. While ideal Airy beams on incoherent background do not accelerate in free space, their soft-apertured realizations, which carry finite power and hence are amenable to the laboratory implementation, have their acceleration strongly suppressed. The actual distance over which any finite-power ABIB defies diffraction is determined by the interplay of the coherence width of the random background and the aperture size. We anticipate the proposed beams to find applications to imaging as well as to optical communications and information transfer through random media. 
 
\

\noindent \textbf{Funding.}~
Natural Sciences and Engineering Research Council of Canada (RGPIN-2018-05497).

\

\noindent \textbf{Disclosures.}~The authors declare no conflicts of interest.

\end{document}